\documentclass{iaus}
\usepackage{graphicx}

\newcommand{\hspx}{\hspace*{1.5cm}}

\title[Evolution of CEMP stars] 
{Modelling the evolution and nucleosynthesis of carbon-enhanced 
metal-poor stars}

\author[Pols et al]   
{O.~R. Pols$^1$, R.~G. Izzard$^1$%
  \thanks{Present address: Institut d'Astronomie et d'Astrophysique, 
  Universit\'e Libre de Bruxelles, CP226, Boulevard du triomphe, 
  B-1050 Bruxelles, Belgium},
M. Lugaro$^1$, S.~E. de Mink$^1$ 
}

\affiliation{$^1$Sterrekundig Instituut Utrecht, P.O. Box 80000, 
NL-3584 TA Utrecht, The Netherlands \break email: O.R.Pols@uu.nl, 
R.G.Izzard@uu.nl, M.A.Lugaro@uu.nl, S.E.deMink@uu.nl}

\pubyear{2008}
\volume{252}  
\date{?? and in revised form ??}
\jname{The Art of Modelling Stars in the 21st Century}
\editors{L. Deng \& C.E. Editor, eds.}
\begin{document}

\maketitle

\begin{abstract} We present the results of binary population simulations
of carbon-enhanced metal-poor (CEMP) stars. We show that nitrogen and
fluorine are useful tracers of the origin of CEMP stars, and conclude
that the observed paucity of very nitrogen-rich stars puts strong
constraints on possible modifications of the initial mass function at
low metallicity.  The large number fraction of CEMP stars may instead
require much more efficient dredge-up from low-metallicity asymptotic
giant branch stars.

\end{abstract}

\firstsection 
\section{Introduction}
One of the most striking results of recent large surveys for very
metal-poor stars in the Galactic halo is the large proportion of
highly carbon-enriched objects among them. These carbon-enhanced
metal-poor (CEMP) stars, usually defined as metal-poor stars with
$\mathrm{[C/Fe]} > 1.0$, make up at least 10 per cent and probably as
much as 20--25 per cent of very metal-poor stars with $\mathrm{[Fe/H]}
< -2$ (\cite[Frebel \etal\ 2006]{2006_Frebel}; \cite[Lucatello \etal\
2006]{2006_Lucatello}).

The majority (about 80 per cent, according to \cite[Aoki \etal\
2007]{2007_Aoki}) of CEMP stars are also enriched in Ba and other
heavy elements produced by slow neutron captures (the $s$-process) in
asymptotic giant branch (AGB) stars.  For these so-called CEMP-s stars
a likely scenario is pollution by mass transfer from a more massive
AGB companion in a binary system, which has since become a white
dwarf. Supporting evidence for this scenario comes from radial
velocity monitoring, which suggests that all CEMP-s stars could
statistically be binaries (\cite[Lucatello \etal\
2005a]{2005_Lucatello_binfrac}).  The remaining fraction of CEMP stars
that do not show s-process enrichments (the CEMP-no stars) are
typically more metal-poor than the CEMP-s stars and so far have shown
no evidence for binarity (\cite[Aoki \etal\ 2007]{2007_Aoki}). These
stars exhibit a variety of abundance patterns, and may have formed
instead from material ejected from rapidly rotating massive stars
(\cite[Meynet \etal\ 2006]{2006_Meynet}) or from faint core-collapse
supernovae (\cite[Umeda \& Nomoto 2005]{2005_Umeda}). Confusing such a
clear distinction are a surprisingly large number of CEMP stars
enhanced in both $s$-process and $r$-process (rapid neutron-capture)
elements, whose origin is still quite unclear (\cite[Jonsell \etal\
2006]{2006_Jonsell}).

Within the mass transfer scenario, the large proportion of CEMP-s
stars requires the existence of a sufficient number of binary systems
with primary components that have undergone AGB nucleosynthesis. In
recent studies (\cite[Lucatello \etal\ 2005b; Komiya \etal\
2007]{2005_Lucatello_IMF,2007_Komiya}) it has been argued that this
requires a different initial mass function (IMF) at low metallicity,
weighted towards intermediate-mass stars. If true, this in turn has
important consequences for the chemical evolution of the halo and, by
implication, of other galaxies. However, the model calculations on
which these estimates are based still contain many uncertainties
regarding the evolution and nucleosynthesis of low-metallicity AGB
stars, the efficiency of mass transfer, and the evolution of the
surface abundances of the CEMP stars themselves. 

In this contribution we explore the effect of some of these
uncertainties on the number fraction of CEMP stars by means of a
binary population synthesis study. Apart from carbon, we concentrate
on nitrogen and fluorine enrichments as possible tracers of the origin
of CEMP stars, the latter motivated by the recent discovery of
fluorine in a CEMP star (see Sect.~2).  In Sect.~3 we present the
first results of our binary population synthesis simulations, and in
Sect.~4 we give our conclusions.

\section{Nitrogen and fluorine in CEMP stars}

Apart from carbon, substantial enhancements of nitrogen with respect
to iron are common among CEMP stars, typically with $\mathrm{[C/N]} >
0$. Detailed AGB nucleosynthesis models of low initial mass ($<
2.5\,M_\odot$) produce carbon (by the 3$\alpha$ reaction during
thermal pulses and subsequent convective dredge-up), but do not
produce nitrogen because it is burned during the same helium shell
flashes. On the other hand, AGB models of higher mass convert the
dredged-up carbon effectively into nitrogen by CN-cycling at the
bottom of the convective envelope (hot bottom burning, HBB). The
surface abundances of these more massive AGB stars approach the
CN-equilibrium ratio of $\mathrm{[C/N]} \approx -2$. Detailed
evolution models of AGB stars (\cite[Karakas \& Lattanzio
2007]{2007_Karakas}) indicate that HBB sets in at significantly lower
mass at low metallicity (between 2.5 and $3\,M_\odot$ at
$\mathrm{[Fe/H]}=-2.3$) than at solar metallicity (around
$5\,M_\odot$). One may thus expect a population of so-called
nitrogen-enhanced metal-poor (NEMP) stars, with $\mathrm{[C/N]} <
-0.5$. Although a few examples of such stars are known, mostly at
$\mathrm{[Fe/H]}<-2.9$, they appear to be very rare (\cite[Johnson
\etal\ 2007]{2007_Johnson}). As we show in Sect.~3 the number of NEMP
stars sets an additional constraint on possible changes to to IMF at
low metallicity.

Recently, \cite{2007_Schuler} derived a super-solar fluorine abundance
of $\mathrm{[F/Fe]} = +2.9$ in the halo CEMP star HE\,1305+0132. This
is the most iron-deficient star, $\mathrm{[Fe/H]} = -2.5$, for which
the fluorine abundance has been measured. Enhancements of carbon and
nitrogen are also measured ($\mathrm{[C/Fe]} = +2.7$; $\mathrm{[N/Fe]}
= +1.6$), and Ba and Sr lines are seen in its spectrum (\cite[Goswami
2005]{2005_Goswami}), placing HE\,1305+0132 in the group of CEMP-s
stars.

\begin{figure}
\includegraphics[width=\textwidth]{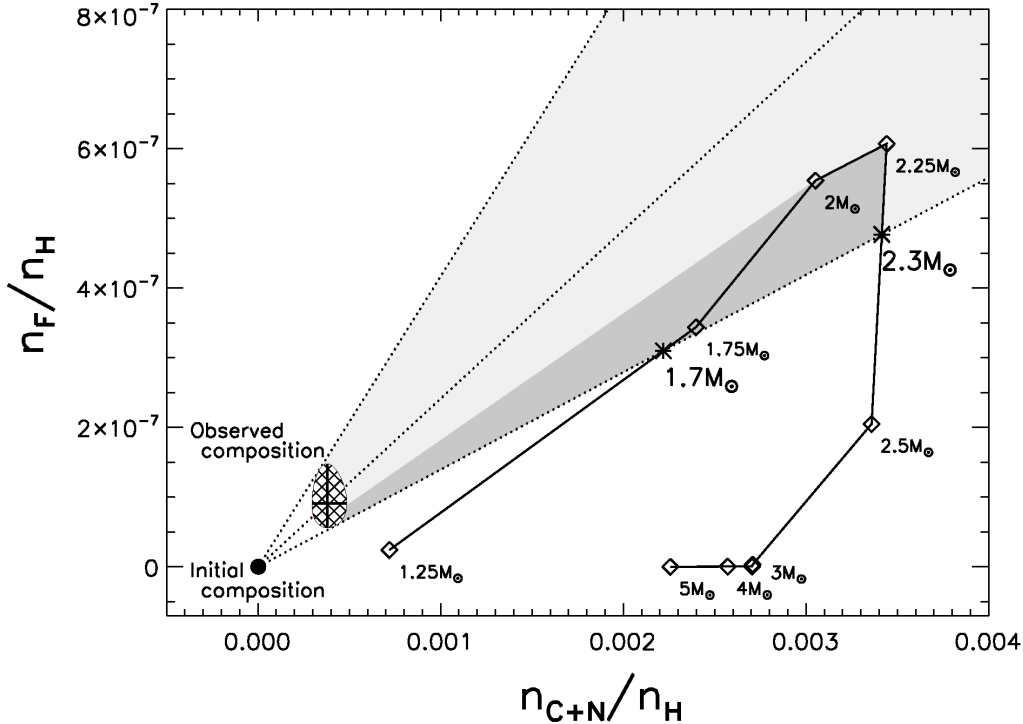}
  \caption{Abundances by number of F and C+N with respect to H as
  observed in HE\,1305+0132 from \cite[Schuler \etal\ (2007; hatched
  ellipsoid showing $1\sigma$ errors)]{2007_Schuler} compared to those
  computed in the material lost in AGB star winds at [Fe/H] $=-2.3$,
  with labels indicating the initial masses. The light gray area is
  the region of the plot where the F and C+N abundances from the AGB
  companion should lie in order to match the observed abundances,
  after dilution. The darker gray area represents the region covered
  by the models with masses in the range $1.7-2.3\,M_\odot$ indicated
  by the asterisks.} \label{fig:HE1305_fluor}
\end{figure}

Fluorine can be made in AGB stars as a by-product of the
$^{14}\mathrm{N} + ^{4}\mathrm{He}$ reaction under neutron-rich
conditions during thermal pulses (\cite[Lugaro \etal\
2004]{2004_Lugaro}). Fluorine enhancements of up to 30 times solar
have been measured among Galactic AGB stars (\cite[Jorissen \etal\
1992]{1992_Jorissen}), demonstrating that these stars can indeed
produce fluorine efficiently. Detailed AGB nucleosynthesis models
(\cite[Karakas \& Lattanzio 2007]{2007_Karakas}) show that fluorine is
produced and dredged to the surface alongside carbon in
low-metallicity AGB stars with masses $< 3\,M_\odot$.  At larger
masses fluorine is destroyed by proton captures as a result of HBB.

In Fig.~\ref{fig:HE1305_fluor} we show the enhancements of F and C+N
relative to hydrogen as observed in HE\,1305+0132 and compare these to
the abundance ratios in the material lost by AGB stars at
$\mathrm{[Fe/H]}=-2.3$ according to the \cite{2007_Karakas} models.
The figure shows that AGB stars with masses between 1.7 and
$2.3\,M_\odot$ produce fluorine and carbon in the right amounts to
account for the observed abundances, after accretion of the material
by a low-mass companion and subsequent dilution in its envelope by a
factor 6 to 9 (\cite[Lugaro \etal\ 2008]{2008_Lugaro}). Assuming a
mass of $0.8\,M_\odot$ and a convective envelope mass fraction of
60\,\%, this implies that the star should have accreted
0.05--$0.12\,M_\odot$ from its companion, corresponding to 3--11\,\%
of the mass lost by the AGB star. These constraints fit well within
the binary mass transfer scenario.  Furthermore this result indicates
the potential of using fluorine as a tracer for the origin of CEMP
stars.

\section{Binary population nucleosynthesis of CEMP stars}

We have simulated populations of metal-poor halo stars in binary
systems using the rapid synthetic binary nucleosynthesis code of
\cite{2004_Izzard} and \cite{2006_Izzard}. The code uses fits to
detailed single-star evolution and nucleosynthesis models, in
particular the AGB models of \cite{2007_Karakas}, and follows the
surface abundances as a star evolves through dredge-up episodes. A
prescription for hot bottom burning in massive AGB stars is included,
calibrated against the same detailed AGB models. Binary evolution is
followed according to the prescriptions of \cite{2002_Hurley}. We
model mass transfer by stellar wind accretion, according to the
\cite{1944_Bondi} prescription, and by Roche-lobe overflow (RLOF). In
binaries with AGB primaries, RLOF is usually unstable and leads to the
ejection of a common envelope without any further accretion onto the
companion. In our default model we assume efficient thermohaline
mixing, in accordance with the findings of \cite{2007_Stancliffe}.

In all models we use solar-scaled initial abundances according to
\cite{1989_Anders} with $Z=10^{-4}$, corresponding to $\mathrm{[Fe/H]}
= -2.3$, which is the lowest metallicity for which we have detailed
AGB models.  We select stars with ages between 10 and
13.7\,Gyr (roughly corresponding to the age of the halo) and $\log g <
4.0$ (thus including turnoff stars and (sub)giants but excluding
unevolved main-sequence stars). Among this sample we designate as CEMP
stars those with $\mathrm{[C/Fe]} > 1.0$ and as NEMP stars those with
$\mathrm{[N/Fe]} > 0.5$ and $\mathrm{[C/N]} < -0.5$, following the
definition of \cite{2007_Johnson}. Note that these definitions partly
overlap.

We can compare our model results with the statistics of the SAGA
database of metal-poor stars (\cite[Suda \etal\ 2008]{2008_Suda}). We
selected 375 stars from the database in a metallicity range
$\mathrm{[Fe/H]}=-2.3\pm0.5$ and $\log g < 4.0$. Of these, 296 have a
C abundance measurement and 69 classify as CEMP stars, yielding a CEMP
fraction of 18--23\,\%. Only one star classifies as a NEMP star,
giving a very small nominal NEMP fraction of $<$0.3\,\%. If we
consider an extended metallicity range $-4<\mathrm{[Fe/H]}<-2$ in
order to improve the number statistics, we find 6 NEMP stars and a
NEMP fraction of about 1.5\,\%. We conclude that CEMP stars outnumber
NEMP stars by at least a factor 10.

\begin{table}
\caption[]{%
   Number fractions of CEMP and NEMP stars among halo stars at
   $\mathrm{[Fe/H]}=-2.3$ and $\log g < 4.0$ resulting from binary
   population synthesis, for different sets of physical ingredients in
   the evolution models. The last two columns give the fraction of
   CEMP-s stars and of fluorine-rich FEMP stars, relative to the
   number of CEMP stars.} \label{tab:results1}
\centerline{%
\begin{tabular}{llccccc}
\hline
model & & $f$(CEMP) & $f$(NEMP) & $\frac{\mathrm{NEMP}}{\mathrm{CEMP}}$ & $\frac{\mathrm{CEMP-s}}{\mathrm{CEMP}}$ & 
  $\frac{\mathrm{FEMP}}{\mathrm{CEMP}}$ \\
\hline
1A & default physics                  & ~2.30\,\% & 0.35\,\% & 0.15 & 0.29 & 0.85 \\
2A & no thermohaline mixing           & ~4.20\,\% & 0.52\,\% & 0.12 & ... & ... \\
3A & calibrated 3DUP                  & ~9.43\,\% & 0.34\,\% & 0.036 & 0.99 & 0.82 \\
4A & no therm.\ mix.\ + calibr.\ 3DUP & 14.96\,\% & 0.47\,\% & 0.031 & 0.75 & 0.87 \\
\hline
\end{tabular}
}
\end{table}

In our default model (1A) we assume that the initial primary masses
$M_1$ are distributed according to the solar neighbourhood IMF as
derived by \cite{1993_Kroupa}, the initial periods come from a flat
distribution in $\log P$ and the initial mass ratios from a flat
distribution in $q=M_2/M_1$. In Table~\ref{tab:results1} we present
the resulting number fractions of CEMP and NEMP stars (first row).
This model clearly fails to account for the large observed CEMP
fraction, although the small NEMP fraction is consistent with the
observations. The last two columns give the number ratio of CEMP-s
stars (defined as those CEMP stars with $\mathrm{[Ba/Fe]} > 0.5$) to
CEMP stars and the ratio of FEMP stars (defined as stars with
$\mathrm{[F/Fe]} > 1.0$) to CEMP stars. Although all our CEMP stars
are also enriched in Ba, in this model the majority have
$\mathrm{[Ba/Fe]} < 0.5$.

In the next rows of Table~\ref{tab:results1} we vary some of the
uncertain physical ingredients in our models, while keeping the input
distributions the same. In model 2A we switch off thermohaline mixing,
which has the effect of increasing the surface abundances of C, N and
Ba in turnoff stars with shallow convection zones relative to our
default model. The numbers of CEMP stars and NEMP stars
correspondingly increase, although the CEMP fraction remains too low
to be compatible with observations. In model 3A we vary some of the
parameters relating to third dredge-up (3DUP), in particular we assume
a smaller value (by up to $0.1\,M_\odot$) of the minimum core mass at
which 3DUP occurs, and after dredge-up has started we assume efficient
3DUP (with $\lambda \geq 0.8$) regardless of how small the envelope
mass is. In practice this means that almost all AGB stars with initial
masses $>0.8\,M_\odot$ undergo efficient dredge-up. These
modifications are motivated by similar changes needed to reproduce the
carbon-star luminosity function of the Magellanic Clouds (\cite[Izzard
\& Tout 2004]{2004_Izzard_CSLF}) and the abundances of Galactic
post-AGB stars (\cite[Bona{\v c}i{\'c} Marinovi{\'c} \etal\
2007]{2007_Bonacic}). This leads to an increase of the CEMP fraction
to almost 10\,\%, much closer to but still short of the observed
value. Model 4A is a combination of 2A and 3A and results in a CEMP
fraction of 15\,\%. We note that in these latter two models nearly all
CEMP stars are CEMP-s stars, in accordance with observations.  We also
note that in all models the majority ($>80$\,\%) of CEMP stars are
expected to be fluorine-rich, with $\mathrm{[F/Fe]} > 1.0$.

\begin{table}
\caption[]{%
   Number fractions of CEMP and NEMP stars among halo stars at
   $\mathrm{[Fe/H]}=-2.3$ and $\log g < 4.0$ as in
   Table~\ref{tab:results1}, for the default physical ingredients
   while varying the input distributions.} \label{tab:results2}
\centerline{%
\begin{tabular}{llccccc}
\hline
model & & $f$(CEMP) & $f$(NEMP) & $\frac{\mathrm{NEMP}}{\mathrm{CEMP}}$ & &  \\
\hline
1A & default $N(M_1,q,P)$                     & ~2.30\,\% & ~0.35\,\% & 0.15 & & \\
1B & $N(q,P)$ from \cite{1991_Duquennoy}      & ~3.50\,\% & ~0.71\,\% & 0.20 & & \\
1C & $N(M_1)$ from \cite{1979_Miller}         & ~3.15\,\% & ~0.62\,\% & 0.20 & & \\
1D & $N(M_1)$ from \cite{2005_Lucatello_IMF}  & ~4.81\,\% & ~1.35\,\% & 0.28 & & \\
1E & $N(M_1)$ from \cite{2007_Komiya}         & 13.47\,\% & 26.61\,\% & 1.98 & & \\
\hline
\end{tabular}
}
\end{table}

In Table~\ref{tab:results2} we present CEMP and NEMP fractions of the
default physical model while varying the initial distributions of
binary parameters. In model 1B we assume mass ratios and periods drawn
from the distributions derived by \cite{1991_Duquennoy} for the local
population of G dwarfs (i.e.\ a log-normal period distribution with a
broad peak at 170 years and a $q$ distribution with a broad peak at
$M_2/M_1=0.23$). This leads to an increase by a factor of 1.5--2 in
the number of CEMP and NEMP stars as the peak in the period
distribution coincides with the period range in which mass transfer is
effective. 

Models 1C, 1D and 1E explore the effect of varying the initial mass
function, by assuming the default $q$ and $P$ distributions in
combination with a log-normal form of the IMF. The \cite{1979_Miller}
IMF also represents the solar neighbourhood but gives somewhat higher
CEMP and NEMP fractions than the \cite{1993_Kroupa} IMF.  Model 1D
assumes the IMF suggested by \cite{2005_Lucatello_IMF} as required to
reproduce the large CEMP fraction (it has a median mass of
$0.79\,M_\odot$, compared to $0.10\,M_\odot$ for the Miller \& Scalo
IMF). It results in a larger CEMP fraction but still falls short of
the observed value. The discrepancy between our and Lucatello's
results arises mainly because in our (default) models the initial
primary mass and period range contributing to CEMP stars are smaller
than they assumed. Model 1D also shows an increased NEMP fraction, the
result of a larger weight of intermediate-mass stars (with $M >
2.7\,M_\odot$ undergoing HBB) in this IMF. This effect is much more
extreme when we assume the IMF suggested by \cite{2007_Komiya} which
has a median mass of $10\,M_\odot$.  Although it gives rise to a
substantial CEMP fraction, the CEMP stars are outnumbered by NEMP
stars by a factor of two. This is not compatible with the observed
limits on the number fraction of NEMP stars.

\section{Conclusions}

The detection of a large fluorine overabundance in the CEMP star
HE\,1305+0132 is well explained within the AGB binary mass transfer
scenario (\cite[Lugaro \etal\ 2008]{2008_Lugaro}). On the other hand,
models of rapidly rotating massive stars do not produce fluorine
(\cite[Meynet \etal\ 2006]{2006_Meynet}). Therefore fluorine appears
to be a useful discriminant between different scenarios proposed for
the origin of CEMP stars.  Our population synthesis results indicate
that (rather independent of model assumptions) at least 80\,\% of CEMP
stars formed by AGB mass transfer should be enriched in fluorine
($\mathrm{[F/Fe]} > 1.0$), i.e.\ most CEMP-s stars should also be FEMP
stars.

Our binary population synthesis models show that the paucity of NEMP
stars among metal-poor halo stars is incompatible with a strongly
modified IMF at low metallicity, heavily weighted towards
intermediate-mass stars, as has been suggested by
\cite{2007_Komiya} in order to explain the high proportion of CEMP
stars. Another possible explanation for the ubiquity of CEMP-s stars
is that low-metallicity AGB stars undergo much more efficient dredge-up
than shown by the detailed evolution models available to date.

\begin{acknowledgments}
We would like to thank Takuma Suda for making the SAGA database of
metal-poor stars available to us in electronic form ahead of
publication.
\end{acknowledgments}

\end{document}